\documentclass[amsmath,amssymb]{revtex4}
\begin{document}
\hspace*{5 in}CUQM-123\\  %
 
\vspace*{0.4 in}

\title{Solutions for certain classes of Riccati differential equation}

\author{Nasser Saad}
\email{nsaad@upei.ca}
\affiliation{Department of Mathematics and Statistics,
University of Prince Edward Island,
550 University Avenue, Charlottetown,
PEI, Canada C1A 4P3.}%Lines break automatically or can be forced with \\

\author{Richard L. Hall}%
 \email{rhall@mathstat.concordia.ca}
\affiliation{Department of Mathematics and Statistics, Concordia University,
1455 de Maisonneuve Boulevard West, Montr\'eal,
Qu\'ebec, Canada H3G 1M8}%

\author{Hakan Ciftci}
 \email{hciftci@gazi.edu.tr}
\affiliation{
Gazi Universitesi, Fen-Edebiyat Fak\"ultesi, Fizik
B\"ol\"um\"u, 06500 Teknikokullar, Ankara, Turkey.
}%

%\date\today% It is always \today, today,

% ----------------------------------------------------------------------
\def\dbox#1{\hbox{\vrule  %  Open box size 2#1 (Abrahams p 273) 
                        \vbox{\hrule \vskip #1
                             \hbox{\hskip #1
                                 \vbox{\hsize=#1}%
                              \hskip #1}%
                         \vskip #1 \hrule}%
                      \vrule}}
\def\qed{\hfill \dbox{0.05true in}}  %  QED
\def\square{\dbox{0.02true in}} % SQUARE 
% ---------------------------------------------------------------------- 

\begin{abstract}
We derive some analytic closed-form solutions for a class of Riccati equations $y'(x)-\lambda_0(x)y(x)\pm y^2(x)=\pm s_0(x)$ where $\lambda_0(x), s_0(x)$ are $C^{\infty}-$functions. We show that if $\delta_n=\lambda_n s_{n-1}-\lambda_{n-1}s_n=0,$
where $\lambda_{n}= \lambda_{n-1}^\prime+s_{n-1}+\lambda_0\lambda_{n-1}\hbox{ and }\quad s_{n}=s_{n-1}^\prime+s_0\lambda_{k-1},\quad n=1,2,\dots,$ then The Riccati equation has a solution given by $y(x)=\mp s_{n-1}(x)/\lambda_{n-1}(x)$. Extension to the generalized Riccati equation $y'(x)+P(x)y(x)+Q(x)y^2(x)=R(x)$  is also investigated.
\end{abstract}

\maketitle

\section{Introduction}

The present authors have recently introduced an iterative technique \cite{Hakan}, known as the asymptotic iteration method (AIM), for the exact and approximate solution of the second-order homogeneous differential equation 
\begin{equation}\label{eq1}
u''(x)=\lambda_0(x)u'(x)+s_0(x)u(x),
\end{equation}
where $\lambda_0(x)$ and $s_0(x)$ are $C^\infty$-differentiable functions.  It was shown that if for sufficiently large $n>0$,
\begin{equation}\label{eq2}
{s_n\over \lambda_n}={s_{n-1}\over \lambda_{n-1}}=\alpha,
\end{equation}
where
\begin{equation}\label{eq3}
\lambda_{n}= \lambda_{n-1}^\prime+s_{n-1}+\lambda_0\lambda_{n-1}\hbox{ ~~and~~ } s_{n}=s_{n-1}^\prime+s_0\lambda_{n-1}, \quad n=1,2,\dots.
\end{equation}
then
\begin{equation}\label{eq4}
y(x)= \exp\left(-\int\limits^{x}\alpha(t) dt\right)
\left[C_2 +C_1\int\limits^{x}\exp\left(\int\limits^{t}(\lambda_0(\tau) + 2\alpha(\tau)) d\tau \right)dt\right]
\end{equation}
is the general solution of the differential equation (\ref{eq1}). Saad et al \cite{saad} proved that the termination condition (\ref{eq2}) is necessary and sufficient for the differential equation (\ref{eq1}) to have polynomial-type solutions. Using the termination condition (\ref{eq2}), the authors were able to show that the classical differential equation of Laguerre, Hermite, Legendre, Jacobi, etc obey this simple criterion. Continuing the work started in \cite{saad}, in the present article we derive some analytic closed-form solutions for different classes of the nonlinear first-order Riccati equation
\begin{equation}\label{eq5}
y'+P(x)y+Q(x)y^2=R(x).
\end{equation}
The principal idea comes from the fact that every Riccati equation reduces to a second order linear differential equation by some suitable transformation \cite{andrei}. Using this idea along with the asymptotic iteration method \cite{Hakan}-\cite{saad}, we construct closed-form solutions for different classes of the Riccati equation (\ref{eq5}) under certain conditions on the functions $P(x)$, $Q(x)$, and $R(x)$. The nonlinear Riccati equation is of great interest for many applications to mathematical physics. It is well known that to find the general solution of the Riccati equation it is enough to know one nontrivial particular solution \cite{andrei}-\cite{in}. 

The article is organized as follows: in the next section we discuss two simple cases $y'-\lambda_0(x)y\pm y^2=\pm s_0(x)$, where $\lambda_0(x)$ and $s_0(x)$ are differentiable functions. In section~3, we discuss the generalized Riccati equation (\ref{eq5}). We show that for certain relations connecting the functions $P(x),Q(x)$ and $R(x),$ we can construct many closed-form solution of the Riccati equation. Selections of results are presented in tables, which can easily be extended.
 
\section{Two simple cases}
\noindent The following Theorem provides condition for the solvability of a certain class of (\ref{eq5}).
\medskip

\noindent{\bf Theorem 1:} \emph{Given $\lambda_0$ and $s_0$ in $C^{\infty}(a,b),$ the Riccati equation 
\begin{equation}\label{eq6}
y'-\lambda_0(x)y+ y^2= s_0(x)
\end{equation}
has a solution 
\begin{equation}\label{eq7}
y_n(x)=-{s_{n-1}(x)\over \lambda_{n-1}(x)}
\end{equation}
if for some $n>0$, $
\delta_{n}=\lambda_{n} s_{n-1}-\lambda_{n-1}s_{n}=0$ 
where $\lambda_n$ and $s_n$ satisfies the recurrence relations (\ref{eq3}).
}
\medskip

The validity of this Theorem can be easily verified by direct substitution of (\ref{eq7}) into (\ref{eq6}) and the application of the termination condition (\ref{eq2}) and the AIM sequence (\ref{eq3}). However, a more constructive proof can be established by substituting $y={u'\over u}$ in Eq.(\ref{eq6}) to obtain the well-known second-order differential equation  (\ref{eq1}). Using the termination condition (\ref{eq2}), we have for $\delta_{n}=\lambda_{n} s_{n-1}-\lambda_{n-1}s_{n}=0$, where $\lambda_n$ and $s_n$ are given by (\ref{eq3}), that the differential equation (\ref{eq1}) has a solution \cite{saad}
\begin{equation}\label{eq8}
u(x)=\exp\bigg(-\int^x{s_{n-1}(\tau)\over \lambda_{n-1}(\tau)}d\tau\bigg),
\end{equation} 
which implies $y(x)={u'(x)\over u(x)}=-{s_{n-1}(x)\over \lambda_{n-1}(x)}$, as required.\qed
\vskip 0.1 true in

\noindent{\bf Example 1:} Consider the Riccati differential equation
\begin{equation*}
y'(x)+{((m-a)x^2+(2cm-1)x-c)\over (ax^3+bx^2+cx)}y(x)+y^2(x)=-{(-2mx+1)\over (ax^3+bx^2+cx)}.
\end{equation*}
With $\lambda_0(x)=-{((m-a)x^2+(2cm-1)x-c)\over (ax^3+bx^2+cx)}$ and $s_0(x)=-{(-2mx+1)\over (ax^3+bx^2+cx)}$, we can easily show $\delta_2=\lambda_2s_1-\lambda_1 s_2=0$, where 
\begin{equation*}
\lambda_2(x)=-{((m+a)x^2+(4cm+2b-1)(x+c))(m(m+3a)x^2+(2m^2c-2(1-b)m-4a)x-3b-3cm+1)\over x(ax^2+bx+c)^3}
\end{equation*} 
and 
\begin{equation*}
s_2(x)={(2(m+a)x+4cm+2b-1)(m(m+3a)x^2+(2m^2c-2(1-b)m-4a)x-3b-3cm+1)\over x(ax^2+bx+c)^3}.
\end{equation*}
Thus a solution is
\begin{equation*}
y(x)={s_1(x)\over \lambda_1(x)}={2(m+a)x+4cm+2b-1\over (m+a)x^2+(4cm+2b-1)(x+c)}.
\end{equation*}

In Table 1, we present some closed-form solutions for different classes of Riccati equation, obtained by direct applications of Theorem 1. In this table, we use the generalized hypergeometric series ${}_pF_q(\alpha_1,\alpha_2,\dots,\alpha_p;\beta_1,\beta_2,\dots,\beta_q;x)$ defined by \cite{prudnikov}
\begin{equation}\label{eq9}
{}_pF_q(\alpha_1,\alpha_2,\dots,\alpha_p;\beta_1,\beta_2,\dots,\beta_q;x)=\sum\limits_{k=0}^\infty {(\alpha_1)_k(\alpha_2)_k \dots(\alpha_p)_k\over (\beta_1)_k(\beta_2)_k\dots(\beta_q)_k}{x^k\over k!}
\end{equation}
where $p$ and $q$ are nonnegative integers and no $\beta_k,k=1,2,\dots,q$ is zero or a negative integer. Clearly, (\ref{eq9}) includes the special cases of the confluent hypergeometric function ${}_1F_1$ and the classical `Gaussian' hypergeometric function ${}_2F_1$. The Pochhammer symbol $(\alpha)_k$ is defined in terms of Gamma function as
\begin{equation}\label{eq10}
(\alpha)_k={\Gamma(\alpha+n)\over \Gamma(\alpha)}\quad k=0,1,2,\dots
\end{equation}
If $\alpha$ is a negative integer $-n$, we have
\begin{equation}\label{eq11}
(-n)_k=\begin{cases} {(-1)^k n!\over (n-k)!}& 0\leq k\leq n\\
0& k>n
\end{cases}
\end{equation}
in which case, the generalized hypergeometric series reduces to a polynomial of degree $n$ in its variable $x$.

\begin{table}[ht]
\caption{Closed-form solutions for the Riccati differential equation $y'-\lambda_0(x)y+ y^2= s_0(x)$ by Theorem 1.} \label{tab:1}
\begin{tabular}{|l|l|}
	\hline
Riccati Equation &  $y_n,~n=0,1,2,\dots$  \\
\hline
$y'-2xy+y^2=-4n$  & $y_n=-4nx{{}_1F_1(-n+1;{3\over 2};x^2)\over {}_1F_1(-n;{1\over 2};x^2)}$\\
\hline
$y'-2xy+y^2=-2(2n+1)$  & $y_n={1\over x}-{4nx\over 3}
{{}_1F_1(-n+1;{5\over 2};x^2)\over {}_1F_1(-n;{3\over 2};x^2)}$\\
\hline
$y'-(ax+b)y+y^2=-2na$  & $y_n=-2n(ax+b){{}_1F_1(-n+1;{3\over 2};{(ax+b)^2\over 2a})\over {}_1F_1(-n;{1\over 2};{(ax+b)^2\over 2a})}$\\
\hline
$y'-(ax+b)y+y^2=-(2n+1)a$  & $y_n={a\over ax+b}-{2n\over 3}(ax+b){{}_1F_1(-n+1;{5\over 2};{(ax+b)^2\over 2a})\over {}_1F_1(-n;{3\over 2};{(ax+b)^2\over 2a})}$\\
\hline
$y'-(b-{c\over x})y+y^2=-{nb\over x}$  & $y_n=-{nb\over c}{{}_1F_1(-n+1;c+1;bx)\over {}_1F_1(-n;c;bx)}$\\
\hline
$y'-{(b-n+1)x-c\over x(1-x)}y+y^2=-{nb\over x(1-x)}$  & $y_n=-{nb\over c}{{}_2F_1(-n+1,b+1;c+1;x)\over {}_2F_1(-n,b;c;x)}$\\
\hline
$y'-{(-2n+1)x-c\over x(1-x)}y+y^2={n^2\over x(1-x)}$  & $y_n={n^2\over c}{{}_2F_1(-n+1,-n+1;c+1;x)\over {}_2F_1(-n,-n;c;x)}$
\\
\hline
$y'-{x\over (1-x^2)}y+y^2=-{n^2\over 1-x^2}$  & $y_n=n^2{{}_2F_1(-n+1,n+1;{3\over 2};{1-x\over 2})\over {}_2F_1(-n,n;{1\over 2};{1-x\over 2})}$\\
	\hline
$y'-{2x\over (1-x^2)}y+y^2=-{n(n+1)\over (1-x^2)}$  & $y_n={n(n+1)\over 2}{{}_2F_1(-n+1,n+2;2;{1-x\over 2})\over {}_2F_1(-n,n+1;1;{1-x\over 2})}$\\
	\hline
$y'-{3x\over (1-x^2)}y+y^2=-{n(n+2)\over (1-x^2)}$  & $y_n={n(n+2)\over 3}{{}_2F_1(-n+1,n+3;{5\over 2};{1-x\over 2})\over {}_2F_1(-n,n+2;{3\over 2};{1-x\over 2})}$\\
	\hline
$y'-{ax\over (1-x^2)}y+y^2=-{n(n+a-1)\over (1-x^2)}$  & $y_n={n(n+a-1)\over a}
{{}_2F_1(-n+1,n+a;{a\over 2}+1;{1-x\over 2})\over {}_2F_1(-n,n+a-1;{a\over 2};{1-x\over 2})}$\\
	\hline
$y'-{(a+b+2)x-b+a\over (1-x^2)}y+y^2=-{n(n+a+b+1)\over (1-x^2)}$  & 
$y_n={n(n+a+b+1)\over 2(a+1)} {{}_2F_1(-n+1,n+a+b+2;a+2;{1-x\over 2})
\over {}_2F_1(-n,n+a+b+1;a+1;{1-x\over 2})}$\\
	\hline
$y'-{(1+2k)x\over (1-x^2)}y+y^2=-{n(n+2k)\over (1-x^2)}$  & 
$y_n={n(n+2k)\over 2}{{}_2F_1(-n+1,n+2k+1;k+{3\over 2};{1-x\over 2})\over {}_2F_1(-n,n+2k;k+{1\over 2};{1-x\over 2})}$\\
	\hline
$y'-{2(1+k)x\over (1-x^2)}y+y^2=-{n(n+2k+1)\over (1-x^2)}$  & 
$y_n={n(n+2k+1)\over 2(k+1)}{{}_2F_1(-n+1,n+2k+2;k+2;{1-x\over 2})\over {}_2F_1(-n,n+2k+1;k+1;{1-x\over 2})}$\\
	\hline
$y'+{2(x+1)\over x^2}y+y^2={n(n+1)\over x^2}$  & 
$y_n={n(n+1)\over 2}{{}_2F_0(-n+1,n+2;-;-{x\over 2})\over {}_2F_0(-n,n+1;-;-{x\over 2})}$\\
	\hline
$y'+{(ax+b)\over x^2}y+y^2={n(n+a-1)\over x^2}$  & 
$y_n={n(n+a-1)\over b}{{}_2F_0(-n+1,n+a;-;-{x\over b})\over {}_2F_0(-n,n+a-1;-;-{x\over b})}$\\
	\hline
\end{tabular}
\end{table}

\vskip0.1in
\noindent{\bf Theorem 2:} \emph{Given $\lambda_0$ and $s_0$ in $C^{\infty}(a,b),$ the Riccati equation 
\begin{equation}\label{eq12}
y'-\lambda_0(x)y- y^2= - s_0(x)
\end{equation}
has a solution 
\begin{equation}\label{eq13}
y_n(x)={s_{n-1}(x)\over \lambda_{n-1}(x)}
\end{equation}
if for some $n>0$, $
\delta_{n}=\lambda_{n} s_{n-1}-\lambda_{n-1}s_{n}=0$ 
where $\lambda_n$ and $s_n$ satisfies the recurrence relations (\ref{eq3}).
}
\vskip0.1true in
\noindent{\bf Proof:~} We substitute $y=-{u'\over u}$ in (\ref{eq12}) and thereby obtain the differential equation (\ref{eq1}) which, by using AIM, yields ${u'\over u}=-{s_{n-1}\over \lambda_{n-1}}$. Therefore the solution of the Riccati equation (\ref{eq12}) is given by (\ref{eq13}).\qed
\vskip0.1true in

\noindent{\bf Example 2:} Consider the Riccati equation
$$y'-(3ax+{1\over x})y-y^2=-a^2,$$
where $a$ is a constant. Direct computation yields
$$\delta_{2n}=\prod_{k=1}^n(a+6k)=0.$$
Consequently, we have
$$y_2=-{2\over x},\quad y_4=-{2(18x^2+1)\over x(1+9x^2)},\quad y_6=-{2(729x^4+108x^2+2)\over x(243x^4+54x^2+2)},\dots $$

In Table~2, we present some closed-form solutions for different classes of Riccati equation, as direct applications of Theorem~2.
 
\begin{table}[ht]
\caption{Closed-form solutions for the Riccati differential equation $y'-\lambda_0(x)y- y^2= - s_0(x)$ by Theorem~2.}  \label{tab:2}
\begin{tabular}{|l|l|}
	\hline
Riccati Equation & $y_n,~n=0,1,2,\dots$\\
\hline
$y'-2xy-y^2=4n$  & $y_n=4nx{{}_1F_1(-n+1;{3\over 2};x^2)\over {}_1F_1(-n;{1\over 2};x^2)}$\\
	\hline
$y'-2xy-y^2=2(2n+1)$  & $y_n=-{1\over x}+{4nx\over 3}
{{}_1F_1(-n+1;{5\over 2};x^2)\over {}_1F_1(-n;{3\over 2};x^2)}$\\
	\hline
$y'-(ax+b)y-y^2=2na$  & $y_n=2n(ax+b){{}_1F_1(-n+1;{3\over 2};{(ax+b)^2\over 2a})\over {}_1F_1(-n;{1\over 2};{(ax+b)^2\over 2a})}$\\
	\hline
$y'-(ax+b)y-y^2=(2n+1)a$  & $y_n=-{a\over ax+b}+{2n\over 3}(ax+b){{}_1F_1(-n+1;{5\over 2};{(ax+b)^2\over 2a})\over {}_1F_1(-n;{3\over 2};{(ax+b)^2\over 2a})}$\\
	\hline
$y'-(b-{c\over x})y-y^2={nb\over x}$  & $y_n={nb\over c}{{}_1F_1(-n+1;c+1;bx)\over {}_1F_1(-n;c;bx)}$\\
	\hline
$y'-{(b-n+1)x-c\over x(1-x)}y-y^2={nb\over x(1-x)}$  & $y_n={nb\over c}{{}_2F_1(-n+1,b+1;c+1;x)\over {}_2F_1(-n,b;c;x)}$\\
	\hline
$y'-{(-2n+1)x-c\over x(1-x)}y-y^2={n^2\over x(1-x)}$  & $y_n=-{n^2\over c} {{}_2F_1(-n+1,-n+1;c+1;x)\over {}_2F_1(-n,-n;c;x)}$\\
	\hline
$y'-{x\over (1-x^2)}y-y^2={n^2\over 1-x^2}$  & $y_n=-n^2{{}_2F_1(-n+1,n+1;{3\over 2};{1-x\over 2})\over {}_2F_1(-n,n;{1\over 2};{1-x\over 2})}$\\
	\hline
$y'-{2x\over (1-x^2)}y-y^2={n(n+1)\over (1-x^2)}$  & $y_n=-{n(n+1)\over 2}{{}_2F_1(-n+1,n+2;2;{1-x\over 2})\over {}_2F_1(-n,n+1;1;{1-x\over 2})}$\\
	\hline
$y'-{3x\over (1-x^2)}y-y^2={n(n+2)\over (1-x^2)}$  & $y_n=-{n(n+2)\over 3}{{}_2F_1(-n+1,n+3;{5\over 2};{1-x\over 2})\over {}_2F_1(-n,n+2;{3\over 2};{1-x\over 2})}$\\
	\hline
$y'-{ax\over (1-x^2)}y-y^2={n(n+a-1)\over (1-x^2)}$  & $y_n=-{n(n+a-1)\over a}
{{}_2F_1(-n+1,n+a;{a\over 2}+1;{1-x\over 2})\over {}_2F_1(-n,n+a-1;{a\over 2};{1-x\over 2})}$\\
	\hline
$y'-{(a+b+2)x-b+a\over (1-x^2)}y-y^2={n(n+a+b+1)\over (1-x^2)}$  & 
$y_n=-{n(n+a+b+1)\over 2(a+1)} {{}_2F_1(-n+1,n+a+b+2;a+2;{1-x\over 2})
\over {}_2F_1(-n,n+a+b+1;a+1;{1-x\over 2})}$\\
	\hline
$y'-{(1+2k)x\over (1-x^2)}y-y^2={n(n+2k)\over (1-x^2)}$  & 
$y_n=-{n(n+2k)\over 2}{{}_2F_1(-n+1,n+2k+1;k+{3\over 2};{1-x\over 2})\over {}_2F_1(-n,n+2k;k+{1\over 2};{1-x\over 2})}$\\
	\hline
$y'-{2(1+k)x\over (1-x^2)}y-y^2={n(n+2k+1)\over (1-x^2)}$  & 
$y_n=-{n(n+2k+1)\over 2(k+1)}{{}_2F_1(-n+1,n+2k+2;k+2;{1-x\over 2})\over {}_2F_1(-n,n+2k+1;k+1;{1-x\over 2})}$\\
	\hline
$y'+{2(x+1)\over x^2}y-y^2=-{n(n+1)\over x^2}$  & 
$y_n=-{n(n+1)\over 2}{{}_2F_0(-n+1,n+2;-;-{x\over 2})\over {}_2F_0(-n,n+1;-;-{x\over 2})}$\\
	\hline
$y'+{(ax+b)\over x^2}y-y^2=-{n(n+a-1)\over x^2}$  & 
$y_n=-{n(n+a-1)\over b}{{}_2F_0(-n+1,n+a;-;-{x\over b})\over {}_2F_0(-n,n+a-1;-;-{x\over b})}$\\
	\hline
\end{tabular}
\end{table}
\medskip
%%%%%%%%%%%%%%%%%%%%%%%%%%%%%%%%%%%%%%%
\section{Generalized Riccati Equation}
%%%%%%%%%%%%%%%%%%%%%%%%%%%%%%%%%%%%%%%
\noindent The differential equation
\begin{equation}\label{eq14}
{dy\over dx}+P(x)y+Q(x)y^2=R(x),
\end{equation} 
is known as the generalized Riccati equation. A number of transformations exist for changing this Riccati equation to a second order linear homogeneous equation (and vice versa). Some of these transformations are summarized in the Table~3.

\vskip0.1in

\noindent{\bf Theorem 3:} \emph{The Riccati equation $y'+P(x)y+Q(x)y^2=R(x)$, has a solution
\begin{equation}\label{eq15}
y(x)=-{s_{n-1}(x)\over Q(x)\lambda_{n-1}(x)}
\end{equation}
where 
\begin{equation}\label{eq16}
\lambda_0={Q'\over Q}-P\quad\text{and}\quad s_0=QR
\end{equation}
and for $n>0$, $\lambda_n$ and $s_n$ are given by $\lambda_{n}= \lambda_{n-1}^\prime+s_{n-1}+\lambda_0\lambda_{n-1}\hbox{ ~~and~~ } s_{n}=s_{n-1}^\prime+s_0\lambda_{n-1}, \quad n=1,2,\dots$.
if for some $n>0$, $\delta_{n}=\lambda_{n} s_{n-1}-\lambda_{n-1}s_{n}=0$.
}
\vskip0.1in
\noindent{\bf Proof:~} We substitute $y={u'\over Qu}$ in the Riccati equation and thereby obtain the second-order differential equation $u''=\big({Q'\over Q}-P\big)u'+QRu$. By using AIM with $\lambda_0={Q'\over Q}-P$ and $s_0=QR$, we have ${u'\over u}=-{s_{n-1}\over \lambda_{n-1}}$ if $\lambda_0$ and $s_0$, along with AIM sequence (\ref{eq3}); thus we satisfy the termination condition $\delta_n=0$. \qed
 
\begin{table}[ht]
\caption{Methods for transforming the generalized Riccati differential equation $y'+P(x)y+Q(x)y^2=R(x)$ to a second order homogeneous differential equation.} \label{tab:3}
\noindent\begin{tabular}{lllll}
\hline
Transformation &~& ~& Resulting Equation&Ref.\\
\hline
$y={u'\over Qu}$ &~& ~& $u''=\bigg({Q'\over Q}-P\bigg)u'+QRu$&\cite{mur}\\
\hline
$y={Ru\over u'}$ &~& ~& $u''=\bigg(P+{R'\over R}\bigg)u'+QRu$&\cite{iwao}\\
\hline
$y={Ru\over u'+Pu}$ &~& ~& $u''=\bigg({R'\over R}-P\bigg)u'+R\bigg(Q-\big({P\over R}\big)'\bigg)u$&\cite{rob}\\
\hline
$y={u'-Pu\over Qu}$ &~& ~& $u''=\bigg(P+{Q'\over Q}\bigg)u'+Q\bigg(\big({P\over Q}\big)'+R\bigg)u$&\cite{rob}\\
\hline
\end{tabular}
\end{table}

\noindent{\bf Example 3:} For the Riccati differential equation
\begin{equation*}
y'(x)+y(x)+e^{{3\over 4}x^4+x}y^2(x)=-27x^2e^{-{3\over 4}x^4-x}.
\end{equation*}
with $\lambda_0(x)={Q'\over Q}-P=3x^3$ and $s_0(x)=QR=-27x^2$, we can easily show $\delta_9=\lambda_9s_8-\lambda_8 s_9=0$. 
Thus a solution of the given differential equation is given by
\begin{equation*}
y(x)=-{s_8(x)\over Q(x)\lambda_8(x)}={9x^8-30x^4+5\over xe^{{3\over 4}x^4+x}(x^8-6x^4+5)}.
\end{equation*}

\noindent{\bf Example 4:}
Consider the Riccati differential equation
\begin{equation*}
y'(x)-{b\over x}y(x)-ax^{n}y^2(x)={c\over x^{n+2}}.
\end{equation*}
Here $Q=-ax^{n}$, $P=-{b\over x}$ and $R={c\over x^{n+2}}$, we have
$\lambda_0(x)={Q'\over Q}-P={n+b\over x}$ and $s_0(x)=QR=-{ac\over x^2}$, we can easily show 
\begin{eqnarray*}
\delta_1&=&\lambda_1s_0-\lambda_0 s_1={ac(-n-b+ac)\over x^4}=0\quad\text{if}\quad ac-(n+b)=0\\
&\Rightarrow& y_1(x)=-{s_0\over Q\lambda_0}=-{1\over ax^{n+1}}\\
\delta_2&=&\lambda_2s_1-\lambda_1 s_2={ac(-n-b+ac)(-2b+2-2n+ac)\over x^6}=0\quad\text{if}\quad ac-2(n+b)+2=0\\
&\Rightarrow& y_2(x)=-{s_1\over Q\lambda_1}=
-{2\over ax^{n+1}}\\
\delta_3&=&\lambda_3s_2-\lambda_2 s_3={ac(-n-b+ac)(-2b+2-2n+ac)(-3b+6-3n+ac)\over x^8}=0\quad\text{if}\quad ac-3(n+b)+6=0\\
&\Rightarrow& y_3(x)=-{s_2\over Q\lambda_2}=
-{3\over ax^{n+1}}\\
\end{eqnarray*} 
and so on. It is clear that $\delta_m=0$ for $m=1,2,\dots$, if $ac-m(n+b)+m(m-1)=0$, and the solution is given by 
\begin{equation*}
y_m(x)=-{m\over a x^{n+1}}\quad\quad\text{for}\quad\quad m=1,2,\dots.
\end{equation*}
This result is expected since the corresponding second-order differential equation is the well-known Cauchy-Euler differential equation.
\vskip 0.1true in
\newpage
\noindent There are some interesting applications that follows from Theorem~3. 

\begin{itemize}
\item If we know that $\lambda_0$ and $s_0$ satisfy the termination condition (\ref{eq2}), we can solve (\ref{eq16}) for $Q(x)$ and $R(x)$ as 
\begin{equation}\label{eq17}
Q(x)=\exp(\int^x(\lambda_0(\tau)+P(\tau))d\tau),\quad R(x)=s_0(x)\exp(-\int^x(\lambda_0(\tau)+P(\tau))d\tau),
\end{equation}
where $P(x)$ is an arbitrary integrable function. Thus, the Riccati equation
\begin{equation}\label{eq18}
y'+P(x)y+e^{\int^x(\lambda_0(\tau)+P(\tau))d\tau}y^2=s_0(x)e^{-\int^x(\lambda_0(\tau)+P(\tau))d\tau}
\end{equation}
has the particular solution
\begin{equation}\label{eq19}
y(x)=-{s_{n-1}(x)\over \lambda_{n-1}(x)}{e^{-\int^x(\lambda_0(\tau)+P(\tau))d\tau}}.
\end{equation}
In order to illustrate this idea, we know \cite{saad} for $\lambda_0=2x$ and $s_0=-2k$ that $\delta_n=0$ for $n=k, k=0,2,4,\dots$. Thus, with $Q(x)=e^{x^2+\int^xP(\tau)d\tau}$ and $R(x)=-4ne^{-x^2-\int^xP(\tau)d\tau}$, the differential equation
$$y'+P(x)y+e^{x^2+\int^xP(\tau)d\tau}y^2=-4ne^{-x^2-\int^xP(\tau)d\tau},$$
has the particular solution
\begin{equation}\label{eq20}
y_n(x)=-4nxe^{-x^2-\int^xP(\tau)d\tau}{{}_1F_1(-n+1;{3\over 2};x^2)\over {}_1F_1(-n;{1\over 2};x^2)}\quad \quad\text{for }\quad\quad n=0,1,2,\dots.
\end{equation}
Furthermore, the differential equation
\begin{equation}\label{eq21}
y'+P(x)y+e^{x^2+\int^xP(\tau)d\tau}y^2=-2(2n+1)e^{-x^2-\int^xP(\tau)d\tau},
\end{equation}
has the solution
\begin{equation}\label{eq22}
y_n=e^{-x^2-\int^xP(\tau)d\tau}\bigg({1\over x}-{4nx\over 3}{{}_1F_1(-n+1;{5\over 2};x^2)\over {}_1F_1(-n;{3\over 2};x^2)}\bigg),\quad\quad n=0,1,2,\dots.
\end{equation}
Note that, in the case of $P(x)=-\lambda_0(x)$, we recover (\ref{eq6}). In Table~4, we present some closed-form solutions for different classes of Riccati equation (\ref{eq18}) for known $\lambda_0(x)$ and $s_0(x)$ \cite{saad}. 
\item The Riccati equation 
\begin{equation}\label{eq23}
y'+\bigg({s_0'(x)\over s_0(x)}-{R'(x)\over R(x)}-\lambda_0(x)\bigg)y+{s_0(x)\over R(x)}y^2=R(x)
\end{equation}
has the particular solution
\begin{equation}\label{eq24}
y(x)=-{R(x)\over s_0(x)}{s_{n-1}(x)\over \lambda_{n-1}(x)}
\end{equation}
if $\lambda_0$ and $s_0$, along with the AIM sequence (\ref{eq3}), satisfies the termination condition $\delta_n=0$.
\item For the Riccati equation  $y'+P(x)y+Q(x)y^2=R(x),$ if
\begin{equation}\label{eq25}
{Q'\over Q}-P+xQR=0,
\end{equation}
then
$y={1\over xQ(x)}$
is a particular solution. For example, the Riccati differential equation
$
y'(x)-xf(x)y-y^2=f(x)
$
for arbitrary differentiable function $f$ has a solution given by $y=-1/x$. This follows from the fact that, if $\lambda_0=-xs_0$, then $\delta_1=0$, and the corresponding second-order differential equation has the solution $u=x$.
\end{itemize}

\begin{table}[ht]
\caption{Closed-form solutions for the Riccati differential equation $y'+P(x)y+ Q(x)y^2=R(x)$ by Theorem~3. Here $P(x)$ is an arbitrary integrable function.} \label{tab:4}
\begin{tabular}{|l|l|l|}
	\hline
$Q(x)$&$R(x)$ &  $y_n,~n=0,1,2,\dots$\\
\hline
$e^{x^2+\int^xP(\tau)d\tau}$&$-4ne^{-x^2-\int^xP(\tau)d\tau}$  & $-4nxe^{-x^2}{{}_1F_1(-n+1;{3\over 2};x^2)\over {}_1F_1(-n;{1\over 2};x^2)}e^{-\int^xP(\tau)d\tau}$\\
\hline
$e^{x^2+\int^xP(\tau)d\tau}$&$-2(2n+1)e^{-x^2-\int^xP(\tau)d\tau}$  & $e^{-x^2}\bigg({1\over x}-{4nx\over 3}{{}_1F_1(-n+1;{5\over 2};x^2)\over {}_1F_1(-n;{3\over 2};x^2)}\bigg)e^{-\int^xP(\tau)d\tau}$\\
\hline
${e^x\over x}e^{\int^x P(\tau)d\tau}$&$-ne^{-x}e^{-\int^x P(\tau)d\tau}$  & $-nxe^{-x}{{}_1F_1(-n+1;2;x)\over {}_1F_1(-n;1;x)}e^{-\int^x P(\tau)d\tau}$\\
\hline
${e^{bx}\over x^c}e^{\int^x P(\tau)d\tau}$&$-nbx^{c-1}e^{-bx}e^{-\int^x P(\tau)d\tau}$&
$-{nb\over c}x^ce^{-bx}{{}_1F_1(-n+1;c+1;bx)\over {}_1F_1(-n;c;bx)}e^{-\int^x P(\tau)d\tau}$\\
\hline
${(x-1)^{c+2n-1}\over x^c}e^{\int^x P(\tau)d\tau}$&${-n^2x^{c-1}\over (x-1)^{2n+c}}e^{-\int^x P(\tau)d\tau}$&$
{n^2x^{c}\over c(x-1)^{2n+c-1}}{{}_2F_1(-n+1,-n+1;c+1;x)\over {}_2F_1(-n,-n;c;x)}e^{-\int^x P(\tau)d\tau}$\\
\hline
${(x-1)^{c+n-b-1}\over x^c}e^{\int^x P(\tau)d\tau}$&${nbx^{c-1}\over (x-1)^{n+c-b}}e^{-\int^x P(\tau)d\tau}$&$
-{nbx^{c}\over c(x-1)^{c+n-1-b}}{{}_2F_1(-n+1,b+1;c+1;x)\over {}_2F_1(-n,b;c;x)}e^{-\int^x P(\tau)d\tau}$\\
\hline
${1\over \sqrt{x^2-1}}e^{\int^x P(\tau)d\tau}$&${n^2\over \sqrt{x^2-1}}e^{-\int^x P(\tau)d\tau}$&$n^2\sqrt{x^2-1}{{}_2F_1(-n+1,n+1;{3\over 2},{1-x\over 2})\over {}_2F_1(-n,n;{1\over 2},{1-x\over 2})}e^{-\int^x P(\tau)d\tau}$\\
\hline
${1\over {x^2-1}}e^{\int^x P(\tau)d\tau}$&$n(n+1)e^{-\int^x P(\tau)d\tau}$&${1\over 2}n(n+1)(x^2-1)e^{-\int^x P(\tau)d\tau}{{}_2F_1(-n+1,n+2;2,{1-x\over 2})\over {}_2F_1(-n,n+1;1,{1-x\over 2})}$\\
\hline
${1\over ({x^2-1})^{3/2}}e^{\int^x P(\tau)d\tau}$&$n(n+2)\sqrt{x^2-1}e^{-\int^x P(\tau)d\tau}$&${n(n+2)\over 3}(x^2-1)^{3/2}e^{-\int^x P(\tau)d\tau}{{}_2F_1(-n+1,n+3;{5\over 2},{1-x\over 2})\over {}_2F_1(-n,n+2;{3\over 2},{1-x\over 2})}$\\
\hline
${1\over ({x^2-1})^{a/2}}e^{\int^x P(\tau)d\tau}$&$n(n+a-1){(x^2-1)^{{a\over 2}-1}}e^{-\int^x P(\tau)d\tau}$&${n(n+a-1)\over a}(x^2-1)^{a/2}e^{-\int^x P(\tau)d\tau}{{}_2F_1(-n+1,n+a;{a\over 2}+1;{1-x\over 2})\over {}_2F_1(-n,n+a-1;{a\over 2};{1-x\over 2})}$\\
\hline
${1\over (x^2-1)^{k+1}}{e^{\int^x P(\tau)d\tau}}$&${n(n+2k+1)}(x^{2}-1)^k e^{-\int^x P(\tau)d\tau}$&${n(n+2k+1)\over 2(k+1)}(x^{2}-1)^{k+1}e^{-\int^x P(\tau)d\tau}{{}_2F_1(-n+1,n+2k+2;k+2;{1-x\over 2})\over {}_2F_1(-n,n+2k+1;k+1;{1-x\over 2})}$\\
\hline
${1\over (x^2-1)^{k+1/2}}{e^{\int^x P(\tau)d\tau}}$&${n(n+2k)}(x^{2}-1)^{k-1/2} e^{-\int^x P(\tau)d\tau}$&${n(n+2k)\over 1+2k}(x^{2}-1)^{k+1/2}e^{-\int^x P(\tau)d\tau}{{}_2F_1(-n+1,n+2k+1;k+{3\over 2};{1-x\over 2})\over {}_2F_1(-n,n+2k;k+{1\over 2};{1-x\over 2})}$\\
\hline
${1\over x^2}e^{2/x}{e^{\int^x P(\tau)d\tau}}$&${n(n+1)}e^{-{2/x}}e^{-\int^x P(\tau)d\tau}$&${n(n+1)\over 2}x^2
e^{-{2/x}}e^{-\int^x P(\tau)d\tau}{{}_2F_0(-n+1,n+2;-;-{x\over 2})\over {}_2F_0(-n,n+1;-;-{x\over 2})}$\\
\hline
${1\over x^a}e^{b/x}{e^{\int^x P(\tau)d\tau}}$&${n(n+a-1)}x^{a-2}e^{-{b/x}}e^{-\int^x P(\tau)d\tau}$&${n(n+a-1)\over b}x^{a}e^{-b/x}e^{-\int^x P(\tau)d\tau}{{}_2F_0(-n+1,n+a;-;-{x\over b})\over {}_2F_0(-n,n+a-1;-;-{x\over b})}$\\
\hline
${1\over (x-1)^{a+1}(x+1)^{b+1}}{e^{\int^x P(\tau)d\tau}}$&${n(n+a+b+1)\over(x-1)^{-a}(x+1)^{-b}}e^{-\int^x P(\tau)d\tau}$&${n(n+a+b+1)(x^2-1)\over 2(a+1)(x-1)^{-a}(x+1)^{-b}}e^{-\int^x P(\tau)d\tau}{{}_2F_1(-n+1,n+a+b+2;a+2;{1-x\over 2})\over {}_2F_1(-n,n+a+b+1;a+1;{1-x\over 2})}$\\
\hline
\end{tabular}
\end{table}
\vskip0.1true in
\noindent By means of the transformation $y={Ru\over u'}$, the following theorem can easily be proved.
\vskip0.1true in

\noindent{\bf Theorem 4:} \emph{The Riccati differential equation  $y'+P(x)y+Q(x)y^2=R(x),$ has the particular solution
\begin{equation}\label{eq26}
y(x)=-{R(x)\lambda_{n-1}(x)\over s_{n-1}(x)},
\end{equation}
if for some $n>0$,
$\delta_n=\lambda_n s_{n-1}-\lambda_{n-1}s_n=0$ where
$\lambda_0={R'\over R}+P$ and $s_0=QR$.}
\vskip0.1in
\noindent In Table 5, we present closed-form solutions for different classes of Riccati equation, obtained as direct applications of Theorem~4.
 
\begin{table}[ht]
\caption{Closed-form solutions for the Riccati differential equation $y'+P(x)y+ Q(x)y^2=R(x)$ by Theorem~4. Here $P(x)$ is an arbitrary integrable function.} \label{tab:5}
\begin{tabular}{|l|l|l|}
	\hline
$Q(x)$&$R(x)$ &  $y_n,~n=1,2,\dots$\\
\hline
$-4ne^{-x^2+\int^xP(\tau)d\tau}$&$e^{x^2-\int^xP(\tau)d\tau}$  & $-{1\over 4nx}e^{x^2}{{}_1F_1(-n;{1\over 2};x^2)\over {}_1F_1(-n+1;{3\over 2};x^2)}e^{-\int^xP(\tau)d\tau}$\\
\hline
$-2(2n+1)e^{-x^2+\int^xP(\tau)d\tau}$&$e^{x^2-\int^xP(\tau)d\tau}$  & ${x~e^{x^2}~{}_1F_1(-n;{3\over 2};x^2)\over {}_1F_1(-n;{3\over 2};x^2)-{4\over 3}x^2n{}_1F_{1}(-n+1;{5\over 2};x^2)}e^{-\int^xP(\tau)d\tau}$\\
\hline
$-2nae^{-{ax^2\over 2}-bx+\int^xP(\tau)d\tau}$&$e^{{ax^2\over 2}+bx-\int^xP(\tau)d\tau}$  & 
$-{e^{{ax^2\over 2}+bx}\over 2n(ax+b)}{{}_1F_1(-n;{1\over 2};{(ax+b)^2\over 2a})\over {}_1F_1(-n+1;{3\over 2};{(ax+b)^2\over 2a})}e^{-\int^xP(\tau)d\tau}$\\
\hline
$-(2n+1)ae^{-{ax^2\over 2}-bx+\int^xP(\tau)d\tau}$&$e^{{ax^2\over 2}+bx-\int^xP(\tau)d\tau}$  & 
${{(ax+b)~e^{{ax^2\over 2}+bx}}~{}_1F_1(-n;{3\over 2};{(ax+b)^2\over 2a})\over a{}_1F_1(-n,{3\over 2},{(ax+b)^2\over 2a})-{2n\over 3}(ax+b)^2{}_1F_1(-n+1;{5\over 2};{(ax+b)^2\over 2a})}e^{-\int^xP(\tau)d\tau}$\\
\hline
$-ne^{-x+\int^xP(\tau)d\tau}$&${1\over x}e^{x-\int^xP(\tau)d\tau}$  & 
$-{1\over  nx}e^{x}{{}_1F_1(-n;1;x)\over{}_1F_1(-n+1;2;x))}e^{-\int^xP(\tau)d\tau}$\\
\hline
$-nbx^{c-1}e^{-bx+\int^xP(\tau)d\tau}$&${1\over x^c}e^{bx-\int^xP(\tau)d\tau}$  & 
${c\over nx^c}e^{bx}{{}_1F_1(-n;c;bx)\over {}_1F_1(-n+1;c+1;bx))}e^{-\int^xP(\tau)d\tau}$\\
\hline
$nbx^{c-1}(x-1)^{(b+1-c-n)}e^{\int^xP(\tau)d\tau}$&$x^{-c}(x-1)^{(c+n-b-1)}e^{-\int^xP(\tau)d\tau}$  & 
${c\over bn}x^{-c}(x-1)^{(c+n-b-1)}{{}_1F_1(-n,b;c;x)\over {}_1F_1(-n+1,b+1;c+1;x))}e^{-\int^xP(\tau)d\tau}$\\
\hline
$-n^2x^{c-1}(x-1)^{(1-c-2n)}e^{\int^xP(\tau)d\tau}$&$x^{-c}(x-1)^{(c+2n-1)}e^{-\int^xP(\tau)d\tau}$  & 
$-{c\over n^2}x^{-c}(x-1)^{(c+2n-1)}{{}_1F_1(-n,-n;c;x)\over {}_1F_1(-n+1,-n+1;c+1;x))}e^{-\int^xP(\tau)d\tau}$\\
\hline
$n(n+1)e^{\int^xP(\tau)d\tau}$&${1\over x^2-1}e^{-\int^xP(\tau)d\tau}$& 
${2\over (x^2-1)}{{}_2F_1(-n,n+1;1;{1-x\over 2})\over{n(n+1)}{}_2F_1(-n+1,n+2;2;{1-x\over 2})}e^{-\int^xP(\tau)d\tau}$\\
\hline
${n(n+a+b+1)\over (x-1)^{-a}(x+1)^{-b}}e^{\int^xP(\tau)d\tau}$&${e^{-\int^xP(\tau)d\tau}\over (x-1)^{a+1}(x+1)^{b+1}}$& 
${(x-1)^{-a-1}\over (x+1)^{b+1}}{2(a+1){}_2F_1(-n,n+a+b+1;a+1;{1-x\over 2})\over{n(n+a+b+1)}{}_2F_1(-n+1,n+a+b+2;a+2;{1-x\over 2})}e^{-\int^xP(\tau)d\tau}$\\
\hline
${n^2\over \sqrt{x^2-1}}e^{\int^xP(\tau)d\tau}$&${e^{-\int^xP(\tau)d\tau}\over \sqrt{x^2-1}}$& 
${e^{-\int^xP(\tau)d\tau}\over n^2\sqrt{x^2-1}}{{}_2F_1(-n,n;{1\over 2};{1-x\over 2})\over{}_2F_1(-n+1,n+1;{3\over 2};{1-x\over 2})}$\\
\hline
$n(n+2)\sqrt{x^2-1}e^{\int^xP(\tau)d\tau}$&${e^{-\int^xP(\tau)d\tau}\over (x^2-1)^{3/2}}$& 
${3e^{-\int^xP(\tau)d\tau}\over n(n+2)(x^2-1)^{3/2}}{{}_2F_1(-n,n+2;{3\over 2};{1-x\over 2})\over{}_2F_1(-n+1,n+3;{5\over 2};{1-x\over 2})}$\\
\hline
$n(n+2k)(x^2-1)^{k-1/2}e^{\int^xP(\tau)d\tau}$&${e^{-\int^xP(\tau)d\tau}\over (x^2-1)^{1/2+k}}$& 
${(1+2k)e^{-\int^xP(\tau)d\tau}\over n(n+2k)(x^2-1)^{k+1/2}}{{}_2F_1(-n,n+2k;k+{1\over 2};{1-x\over 2})\over{}_2F_1(-n+1,n+2k+1;k+{3\over 2};{1-x\over 2})}$\\
\hline
$n(n+2k+1)(x^2-1)^{k}e^{\int^xP(\tau)d\tau}$&${e^{-\int^xP(\tau)d\tau}\over (x^2-1)^{1+k}}$& 
${2(1+k)e^{-\int^xP(\tau)d\tau}\over n(n+2k+1)(x^2-1)^{k+1}}{{}_2F_1(-n,n+2k+1;k+1;{1-x\over 2})\over{}_2F_1(-n+1,n+2k+2;k+2;{1-x\over 2})}$\\
\hline
$n(n+1)e^{-{2\over x}+\int^xP(\tau)d\tau}$&${1\over x^2}{e^{{2\over x}-\int^xP(\tau)d\tau}}$& 
${2e^{{2\over x}-\int^xP(\tau)d\tau}\over n(n+1)x^2}{{}_2F_0(-n,n+1;-;-{x\over 2})\over{}_2F_0(-n+1,n+2;-;-{x\over 2})}$\\
\hline
${n(n+a-1)\over x^{2-a}}e^{-{b\over x}+\int^xP(\tau)d\tau}$&${1\over x^a}{e^{{b\over x}-\int^xP(\tau)d\tau}}$& 
${be^{{b\over x}-\int^xP(\tau)d\tau}\over n(n+a-1)x^a}{{}_2F_0(-n,n+a-1;-;-{x\over b})\over{}_2F_0(-n+1,n+a;-;-{x\over 2})}$\\
\hline
\end{tabular}
\end{table}
\vskip0.1true in

\noindent By means of the transformation $y={Ru\over u'+Pu}$, it it becomes straightforward to prove the following theorem.

\vskip0.1true in

\noindent{\bf Theorem 5:} \emph{The Riccati differential equation  $y'+P(x)y+Q(x)y^2=R(x),$ has the particular solution
\begin{equation}\label{eq27}
y(x)={R(x)\lambda_{n-1}(x)\over -s_{n-1}(x)+P(x)\lambda_{n-1}(x)},
\end{equation}
if for some $n>0$,
$\delta_n=\lambda_n s_{n-1}-\lambda_{n-1}s_n=0$ where
$\lambda_0={R'\over R}-P$ and $s_0=R\big(Q-\big({P\over R}\big)'\big)$.}
\vskip0.1in
\noindent There are two immediate consequence of this Theorem:
\begin{itemize}
\item For an arbitrary function $R(x)$, the Riccati equation
\begin{equation}\label{eq28}
y'+\bigg({R'(x)\over R(x)}-\lambda_0(x)\bigg)y+\bigg({s_0(x)\over R(x)}+\bigg({R'(x)\over R^2(x)}-{\lambda_0(x)\over R(x)}\bigg)'\bigg)y^2=R(x)
\end{equation}
has the particular solution
\begin{equation}\label{eq29}
y_n(x)={R(x)\lambda_{n-1}(x)\over -s_{n-1}(x)+\bigg({R'(x)\over R(x)}-\lambda_0(x)\bigg)\lambda_{n-1}(x)}\quad\quad\quad\text{for}\quad n=0,1,2,\dots
\end{equation}
if $\lambda_0(x)$ and $s_0(x)$ satisfy the termination condition (\ref{eq2}), namely $\delta_n=\lambda_ns_{n-1}-\lambda_{n-1}s_n=0$ for $n=0,1,2,\dots$. Note that, the iteration sequence (\ref{eq3}) can start with $n=0$ for $\lambda_{-1}=1$ and $s_{-1}=0$. In Table~6, we exhibit closed-form solutions for the Riccati equation (\ref{eq14}).
 
\item For an arbitrary function $P(x)$, the Riccati equation
\begin{equation}\label{eq30}
y'
+P(x)y
+\bigg[s_0(x)e^{-\int^x(\lambda_0(\tau)+P(\tau))d\tau}+\bigg(P(x)e^{-\int^x(\lambda_0(\tau)+P(\tau))d\tau}\bigg)'\bigg]y^2
=e^{\int^x(\lambda_0(\tau)+P(\tau))d\tau}
\end{equation}
has the particular solution
\begin{equation}\label{eq31}
y_n={\lambda_{n-1}(x)e^{\int^x(\lambda_0(\tau)+P(\tau))d\tau}\over -s_{n-1}(x)+P(x)\lambda_{n-1}(x)}\quad\quad\quad\text{for}\quad n=0,1,2,\dots
\end{equation}
if $\lambda_0(x)$ and $s_0(x)$ satisfy the termination condition (\ref{eq2}), namely $\delta_n=\lambda_ns_{n-1}-\lambda_{n-1}s_n=0$ for $n=0,1,2,\dots$. 
\end{itemize}

\begin{table}[ht]
\caption{Closed-form solutions for the Riccati differential equation $y'+P(x)y+ Q(x)y^2=R(x)$ by Theorem~5. Here $R(x)$ is an arbitrary differentiable function.} \label{tab:6}
\begin{tabular}{|l|l|}
	\hline
Riccati Equation &  Solution $y_n$, $n=0,1,2,\dots$\\
\hline
$y'+\bigg({R'(x)\over R(x)}-2x\bigg)y+\bigg(-{4n\over R(x)}+\bigg({R'(x)\over R^2(x)}-{2x\over R(x)}\bigg)'\bigg)y^2=R(x)
$& 
${R(x)\over -4nx~{{}_1F_1(-n+1;{3\over 2};x^2)
\over {}_1F_1(-n;{1\over 2};x^2)}
+{R'(x)\over R(x)}-2x}$\\
\hline
$y'+\bigg({R'(x)\over R(x)}-2x\bigg)y+\bigg(-{2(2n+1)\over R(x)}+\bigg({R'(x)\over R^2(x)}-{2x\over R(x)}\bigg)'\bigg)y^2=R(x)
$& 
${R(x)\over -{1\over x}+{4nx\over 3}{{}_1F_1(-n+1;{5\over 2};x^2)\over
{}_1F_1(-n;{3\over 2};x^2)}+{R'(x)\over R(x)}-2x}$\\
\hline
$y'+\bigg({R'(x)\over R(x)}-ax-b\bigg)y+\bigg(-{2na\over R(x)}+\bigg({R'(x)\over R^2(x)}-{(ax+b)\over R(x)}\bigg)'\bigg)y^2=R(x)
$& 
${R(x)\over -2n(ax+b){{}_1F_1(-n+1;{3\over 2};{(ax+b)^2\over 2a})\over
{}_1F_1(-n;{1\over 2};{(ax+b)^2\over 2a})}+{R'(x)\over R(x)}-ax-b}$\\
\hline
$y'+\bigg({R'(x)\over R(x)}-ax-b\bigg)y+\bigg(-{(2n+1)a\over R(x)}+\bigg({R'(x)\over R^2(x)}-{(ax+b)\over R(x)}\bigg)'\bigg)y^2=R(x)
$& 
${R(x)\over -{a\over ax+b}+{2n\over 3}{(ax+b){}_1F_1(-n+1;{5\over 2};{(ax+b)^2\over 2a})
\over {}_1F_1(-n;{3\over 2};{(ax+b)^2\over 2a})}
+{R'(x)\over R(x)}-ax-b}$\\
\hline
$y'+\bigg({R'(x)\over R(x)}-b+{c\over x}\bigg)y+\bigg(-{nb\over xR(x)}+\bigg({R'(x)\over R^2(x)}-{(b-{c\over x})\over R(x)}\bigg)'\bigg)y^2=R(x)
$& 
${R(x)\over -{{nb}~{}_1F_1(-n+1;{c+1};bx)\over c~{}_1F_1(-n;c;bx)}
+{R'(x)\over R(x)}-b+{c\over x}}$\\
\hline
$y'+\bigg({R'(x)\over R(x)}-{(-2n+1)x-c\over x(1-x)}\bigg)y+
\bigg({n^2\over x(1-x)R(x)}+\bigg({R'(x)\over R^2(x)}-{{(-2n+1)x-c}\over x(1-x) R(x)}\bigg)'\bigg)y^2=R(x)
$& 
${R(x)\over {{n^2}~{}_2F_1(-n+1,-n+1;{c+1},x)\over c~~{}_2F_1(-n,-n;c;x)}
+{R'(x)\over R(x)}-{(-2n+1)x-c\over x(1-x)}}$\\
\hline
$y'+\bigg({R'(x)\over R(x)}-{(-n+b+1)x-c\over x(1-x)}\bigg)y+
\bigg({-nb\over x(1-x)R(x)}+\bigg({R'(x)\over R^2(x)}-{{(-n+b+1)x-c}\over x(1-x) R(x)}\bigg)'\bigg)y^2=R(x)
$& 
${R(x)\over {{-nb}~{}_2F_1(-n+1,b+1;{c+1},x)\over c~~{}_2F_1(-n,-n;c;x)}
+{R'(x)\over R(x)}-{(-n+b+1)x-c\over x(1-x)}}$\\
\hline
$y'+\bigg({R'(x)\over R(x)}+{(ax+b)\over x^2}\bigg)y+
\bigg({n(n+a-1)\over x^2R(x)}+\bigg({R'(x)\over R^2(x)}+{{(ax+b)}\over x^2R(x)}\bigg)'\bigg)y^2=R(x)
$& 
${R(x)\over {n(n+a-1)~{}_2F_0(-n+1,n+a;-;-{x\over b})\over b~~{}_2F_0(-n,n+a-1;-;-{x\over b})}
+{R'(x)\over R(x)}+{(ax+b)\over x^2}}$\\
\hline
$y'+\bigg({R'(x)\over R(x)}+{2(k+1)x\over (1-x^2)}\bigg)y+
\bigg(-{n(n+2k+1)\over (1-x^2)R(x)}+\bigg({R'(x)\over R^2(x)}-{{2(k+1)x}\over (1-x^2)R(x)}\bigg)'\bigg)y^2=R(x)
$& 
${R(x)\over {n(n+2k+1)~{}_2F_1(-n+1,n+2k+2;k+2;{1-x\over 2})\over 2(k+1)~{}_2F_1(-n,n+2k+1;k+1;{1-x\over 2})}
+{R'(x)\over R(x)}-{2(k+1)x\over (1-x^2)}}$\\
\hline
$y'+\bigg({R'(x)\over R(x)}+{(2k+1)x\over (1-x^2)}\bigg)y+
\bigg(-{n(n+2k)\over (1-x^2)R(x)}+\bigg({R'(x)\over R^2(x)}-{{(2k+1)x}\over (1-x^2)R(x)}\bigg)'\bigg)y^2=R(x)
$& 
${R(x)\over {n(n+2k)~{}_2F_1(-n+1,n+2k+1;k+{3\over 2};{1-x\over 2})\over (2k+1)~{}_2F_1(-n,n+2k;k+{1\over 2};{1-x\over 2})}
+{R'(x)\over R(x)}-{(2k+1)x\over (1-x^2)}}$\\
\hline
\end{tabular}
\end{table}
\vskip0.1 true in
\noindent By means of the transformation $y={u'-Pu\over Qu}$, it then becomes strarightforward to prove the following theorem.
\vskip0.1 true in
\noindent{\bf Theorem 6:} \emph{The Riccati differential equation  
${dy\over dx}+P(x)y+Q(x)y^2=R(x),$ 
has the particular solution
\begin{equation}\label{eq32}
y(x)={-s_{n-1}(x)-P(x)\lambda_{n-1}(x)\over Q(x)\lambda_{n-1}(x)},
\end{equation}
if for some $n>0$,
$\delta_n=\lambda_n s_{n-1}-\lambda_{n-1}s_n=0$ where
$\lambda_0=P+{Q'\over Q}$ and $s_0=Q\big(\big({P\over Q}\big)'+R\big)$.
}
\vskip0.1true in
\noindent An immediate result implied by this theorem is the following: for arbitrary $P(x)$, the Riccati equation
\begin{equation}\label{eq33}
y'+P(x)y+e^{\int^x(\lambda_0(\tau)-P(\tau))d\tau}y^2=s_0(x)e^{-\int^x(\lambda_0(\tau)-P(\tau))d\tau}-\bigg(P(x)e^{-\int^x(\lambda_0(\tau)-P(\tau))d\tau}\bigg)'
\end{equation}
has the particular solution
\begin{equation}\label{eq34}
y(x)=\bigg[-{s_{n-1}(x)\over \lambda_{n-1}(x)}\bigg]e^{-\int^x(\lambda_0(\tau)-P(\tau))d\tau}
\end{equation}
%%%%%%%%%%%%%%%%%%%%%%%%%%%%%%%%%%%%%%%
\section{Conclusion}
%%%%%%%%%%%%%%%%%%%%%%%%%%%%%%%%%%%%%%%
\noindent It is well know that a Riccati equation can be transformed to a second-order linear differential equation by means of a suitable transformation. Using this fact along with a criterion, recently introduced, which guarantees the existence of polynomial solutions to second-order linear differential equations, we are able to derive analytic closed-form solutions for different classes of Riccati equation. By using the methods developed in this paper, the tables of solutions we present can easily extended. For any given pair of differentiable functions, $\lambda_0$ and $s_0$, satisfying the termination condition (\ref{eq2}) along with (\ref{eq3}), a corresponding class of exactly solvable Riccati equation can be generated.  

% ------------------------------------------------------   
\section*{Acknowledgments}
% ------------------------------------------------------
\medskip
\noindent Partial financial support of this work under Grant Nos. GP3438 and GP249507 from the 
Natural Sciences and Engineering Research Council of Canada is gratefully 
acknowledged by two of us (respectively [RLH] and [NS]).

\end{document}